\documentstyle[12pt]{article}

\author{O.Z. Alekperov\thanks{e-mail: semic@lan.ab.az}\\
Institute of Physics of Academy of Sciences of\\
Azerbaijan Republic, 370143, Baku, G. Javid \\
avenue 33}
\title{$n-GaAs$ QUALITY DIAGNOSE FROM SHALLOW IMPURITIES PHOTOELECTRIC\\
SPECTROSCOPY LINE SHAPES DEPENDENCE ON ELECTRIC FIELD
}
\date{ 
}
\input tcilatex

\QQQ{Last Revised}{
}

\QQQ{Language}{
American English
}

\begin{document}

\maketitle
\begin{abstract}
It is established experimentally that the low temperature photoelectric
spectra line width of shallow impurities depends not only on charged
impurity concentration $N_i=2KN_A$ and degree of samples compensation $%
K=N_A/N_D$, as it was believed earlier.To a great extent it depends on the
impurity distribution inhomogeneity also.For samples with homogeneous and
inhomogeneous distribution of impurities line width dependence character on
external electric fields, smaller than break down one, are different.This
broadening mechanism allows to control the quality of samples with nearly
equal impurity concentrations.
\end{abstract}

\section{INTRODUCTION}

Shallow impurities photoelectric spectroscopy (SIPS) of hydrogen like
impurities consists in arising of lines in photoconductivity spectra of
semiconductor, when electrons localized on impurities are photo excited
resonantly. It is an effective method for investigations of SI in
semiconductors[1]. Considerable information can be obtained from the
investigations of the line broadening mechanisms [2-4] and experimental line
width. Convenient for this purpose are optically allowed transition lines
from the ground state of SI $(1s)$ to excited Zeeman states $2p_{-1}$ and $%
2p_{+1}$ formed in magnetic field $H$. Even for available purest samples of
Si, Ge, GaAs, InP the low temperature SI photoexcitation lines dominant
broadening mechanism is concentration mechanism (inhomogeneous broadening).
This broadening is the result of optical transition (for example $%
1s\rightarrow 2p_{+1}$) energy differences for all impurity atoms from that
of isolated impurity atom owing to interaction with surrounding impurities.
In particular, in compensated semiconductors ($K\geq 01$) with charged
impurities concentration $N_i=2KN_D$ comparable with $N_{D\text{ }}$at
moderate magnetic fields ($\gamma =h\omega _c/2Ry^{*}\sim 1$, $h\omega _c$%
-cyclotron energy, $Ry^{*}$- effective Ridberg ), when quadrupole-electric
field gradient broadening for $1s\rightarrow 2p_{\pm 1}$ lines is
negligible[2 ], the broadening of $1s\rightarrow 2p_{+1}$ line arise from
the quadratic Stark effect shift of levels by charged impurities electric
fields $E$. Then the line shape would be determined by distribution of $E^2$
at the neutral donor center, which depends on charged impurity distribution
in semiconductor.

Character of distribution of charged impurity around the optically active
neutral donor depends on two factors. The first is the distribution of
impurity atoms in semiconductor, which mainly is determined with growing
technology (we consider the residual impurities only). The inhomogeneous in
impurity distribution creates the random potential which is the potential of
fluctuations. The influence of this potential on cyclotron resonance line
shape was established [5] and was used for quality control of semiconductor
samples with nearly equal neutral and charged impurity concentrations
[6].The second is how the electrons are distributed on impurities. In the
case of high temperature limit $(T=\infty )$ it is a random distribution.
But in the zero temperature limit the distribution is correlated, which
provides the Coulomb interaction energy of system of charged acceptors,
donors and electrons to be minimized. In the random electron distribution
case the $1s\rightarrow 2p_{+1}$ transition line width depends on charged
impurity concentration $N_i$ only, independently on sample compensation $%
K=N_A/N_D$ [2-3]. In the correlated distribution case acceptors are charged
taking electrons from nearest donor. Then with the same value of $N_i$ the
line width is smaller and strongly depends on $K$.

Distribution transforms from the correlated to the random at temperatures
[7] $k_BT\gg e^2/(\epsilon _0\cdot r_m)$ ($r_m=(4\pi /3N_i)^{1/3}$ is the
mean distance between charged impurities, $\epsilon _0$- static dielectric
constant, $k_B$ - the Boltzman constant). In the case of strong
inhomogeneous impurity system they are collected into the clusters, which
have higher ionized impurity concentration (because of pure regions are
between them) than that of mean value for all sample, determined from Hall
measurements. Note that only these clusters are optically active for
impurity absorption of radiation. So, strong inhomogeneous in the impurity
distribution must cause an additional broadening mechanism for impurity
absorption lines.

Note that the distribution can be changed from correlated to random not with
heating only , but under any external perturbation increasing the energy,
for example, an electric field.

This work devoted to experimental investigations of the SIPS line broadening
caused by SI inhomogeneous distribution in $n$-GaAs epitaxial layers
obtained by liquid and gas phase epitaxy.

\section{EXPERIMENT}

We investigate the $1s\rightarrow 2p_{+1}$ transition line shape of residual
donors in $n-GaAs$ using the laser magnetic photoelectric spectroscopy
method at moderate magnetic fields up to 6.5 T. Photoconductivity spectra
(PS) of samples were registered at fixed far infrared $H_2O$ and $D_2O$
-lasers radiation at wavelengths $\sim 119$ and $\sim 84mcm$ as a function
of magnetic field in the intervals $(3.5\div 3.7)T$ and $(6.0\div 6.2)T$
correspondingly. PS were measured at different electric fields- from linear
region of samples CVC up to SI break down electric field values. At electric
fields smaller than the break down one the constant voltage regime $%
(R_{load}\ll R_{sample})$ and at the break down the constant current regime $%
(R_{sample}\ll R_{load})$ was used. In order the true line shape to be
registered the precaution was taken the PC, which is proportional to the
change of sample conductivity $\bigtriangleup \sigma $, to be linear to
radiation intensity in both regimes. Parameters of samples and growth
technology are given in table1.Samples of $n-GaAs$ obtained by liquid phase
(LP) and gas phase (GP) growth technology were used. The PC measurements
were carried out by cross-modulation method with modulation frequency of
radiation intensity $750$ $Hz$.

For samples investigated the random distribution of electrons takes place at
temperatures $T\gg 20K$. So we suppose that at measurements temperature $%
T=4.2K$ in the linear region of CVC the distribution is nearly correlated.

\section{RESULTS\ AND\ DISCUSSIONS}

The results of $1s\rightarrow 2p_{+1}$ line shape investigations for
different n-GaAs samples are given in table. The width $\triangle E_{}$ was
measured at the half height of the line in magnetic unites $\triangle H$ and
then it was transformed into energetic unites $\triangle E=(\partial
E_{1s\rightarrow 2p_{+1}}/\partial H)\triangle H$. The rapidity of $%
1s\rightarrow 2p_{+1}$ transition energy increase in magnetic field $%
\partial E_{1s\rightarrow 2p_{+1}}/\partial H\approx 0.18$ $meV/kOe$ was
determined from the two magnetic field positions (as indicated in fig 1) of
this line at different laser quantum energy $\epsilon _{119}=10.45meV$ and $%
\epsilon _{84}=14.71meV$.

First we will analyze the $1s\rightarrow 2p_{+1}$ line shape at small
electric fields far from the break down one.The main results are given below.

As seen from table it is hard (except the purest sample7) to find any
correlation between the line width and SI\ concentration, on the one hand,
and degree of compensation on the other hand, in the intervals $N_D\sim
10^{14}-10^{15}$cm$^{-3}$, $K\sim 0.2-0.9$. As a rule samples obtained by LP
technology have relatively broader SI $1s\rightarrow 2p_{+1}$
photoexcitation lines than those obtained by GP technology. For example in
spite of higher charged impurity concentration and compensation degree in
sample1, sample5 has almost twice broader $1s\rightarrow 2p_{+1}$ donors
lines in the linear CVC region. This fact can be explained only by more
inhomogeneous distribution of SI in LP samples. The quantities $N_D$ and $K,$
usually determined from Hall measurements temperature dependency,are mean
values for all sample volume. When impurities are distributed homogeneously
then $N_D$ and $K$ values in any part of sample equal to the mean values
obtained from Hall measurements given in table1.In the strong inhomogeneouse
distribution case in the optically active parts of sample, where absorption
of radiation takes place, the impurities form clusters with higher
concentration than its mean value. It is obvious that this will cause to
broaden the lines.

Except the purest, all samples have symmetric line shape as it is shown in
figure1 for two of them. Only very high purity sample's line shape is
asymmetric (figure2) , having nearly all width at lower energy (or higher
magnetic field) side, as it must be for quadratic Stark effect broadening
mechanism.

It is also notable that the line width of all samples (except the purest)
with $N_D\geq 10^{14}cm^{-3}$ are not depend on magnetic field value as it
can be seen from the comparison of PS at different magnetic fields for two
samples shown in figure 1.But for both knowing inhomogeneouse (quadratic
Stark and quadrupole-gradient) broadening mechanisms in the indicated
regions of magnetic fields the $1s\rightarrow 2p_{+1}$ line width must
decrease with increasing of magnetic field. Narrowing of $1s\rightarrow
2p_{+1}$ donor's lines with magnetic field increasing was observed in very
pure n-GaAs samples[8].

All above given experimental results witnesses that we deal with an
additional inhomogeneouse SI line broadening mechanism connected with
inhomogeneouse distribution of impurities. The energy of fluctuations
increases with increasing of impurity concentration and compensation degree
as $E_0\sim 0.3e^2N_D^{1/3}K^{1/2}\epsilon _0$ [7] and in n- GaAs at
concentrations $N_D>10^{14}cm^{-3}$ it becomes SIPS lines dominant
broadening mechanism. Note that for samples used in table1 the value of $%
E_0\sim 0.1-0.2$ meV is about experimental line width. Because of this
potential depends not only on impurity concentration, but on distribution
character of SI also, this broadening mechanism can be used in quality
control of samples with nearly $N_D$ and $K.$

This conclusion is in accordance with the results of our investigations of
line width dependence on electric field. It was found that at electric
fields smaller than break down the dependence of $1s\rightarrow 2p_{+1}$
SIPS line width on electric field is different for different samples.In
figure 3 this dependence is shown for two samples with inhomogeneouse and
homogeneous SI distribution. With increasing of applied electric field
smaller than break down, we observed both- increasing as well as decreasing
line width (curves 1 and 2 in fig3). At ''candle''like break down region of
CVC for all samples drastic narrowing of $1s\rightarrow 2p_{+1}$ SIPS lines
takes place [9]. The narrowing at break down is the result of screening of
charged impurity Coulomb potential by free electrons arising in result of
theirs avalanche throwing from the part neutral donors into the conduction
band.

If consider electric fields smaller than break down, the experiments show
that only in inhomogeneous samples we have a gradual narrowing of $%
1s\rightarrow 2p_{+1}$ SIPS lines with increasing of electric field, as it
indicated in fig.3 (curve 1).This can be explained in the following way. Far
wings of line in PS correspond to those neutral donors which are placed at
high potential fluctuations and as a consequence around them the charged
impurity concentration is higher than the mean value for sample. External
voltage applied on inhomogeneous sample will distributed inhomogeneously
creating more electric field values in parts of sample with higher charged
impurity concentration. On the other hand in clusters with high impurity
concentration the potential of ionization must be lowered (for example in
Mott transition case, when $N_D\cdot a_B^{*}\geq 0.25$, for n-GaAs taking
place at $N_D>2\cdot 10^{16}cm^{-3}$ the SI break down potential is zero
because all donor electrons already are free). For this reason with
increasing of electric field the neutral donors in clusters with high
impurity concentration (which cause the PC signal far from the line center)
must to be ionized the first. In turn this would cause narrowing of $%
1s\rightarrow 2p_{+1}$ line.

For samples with homogeneous SI distribution at fields $E<E_{bd}$ the $%
1s\rightarrow 2p_{+1}$ line width increase with electric field.The typical $%
1s\rightarrow 2p_{+1}$ line width dependence on applied voltage for GP
samples is shown in fig.3 (curve1). Our comparison of theories for
uncorrelated [3] $(T=\infty )$ and correlated [4] $(T=0)$ electron
distributions the quadratic Stark broadening calculations for GaAs sample
with $N_D=4\cdot 10^{14}$cm$^{-3}$and $K=0.5$ shows that in the first case
the charged impurities electric field distribution function $P(E^2)$ gives
the $1s\rightarrow 2p_{+1}$ line an order of magnitude (20 times) broader.
In the first case we take $P(E^2)=\frac 2\pi E^2\stackrel{\infty }{%
\stackunder{0}{\int }}dx\cdot x\cdot \exp (-\beta x^{\frac 32})\cdot \sin
(E^2x)$ after the analogy of Holtzmark distribution, where $E$- electric
field in units of $eN_{i\text{ }}^{2/3}/\epsilon _0$, and $\beta =\frac{14}{%
15}(2\pi )^{3/2}$. So, we believe that at $T=4.2K$ an external electric
field changes the electron distribution towards to uncorrelated, so that to
explain the 1.5 times broadening of the $1s\rightarrow 2p_{+1}$ line, as it
was observed for GP sample1. Note that as it is seen from curve 1 of fig. 3
strong increase of line width takes place near the break down field. Note
that near the break down field electron distribution changes must be strong.

Another experiment directly shows that the external electric field smaller
than $E_{bd}$ considerably changes the distribution of electrons on donors.
If, there are different kinds of SI in semiconductor with comparable
concentrations, as it is shown in PS of samples1, 6 and 7, the real ground
state would be the $1s$ state of donor having maximal chemical shift value
(the first line of PS in magnetic field). As a result of this, at low
temperature, when electron distribution is correlated, the acceptors must,
basically, to be charged on account of donors having minimum central cell
correction (chemical shift) value (the last line of PS in magnetic field).
In fig.4 the dependence on electric field is shown for the relation of
intensities of these lines $(S_1/S_2)$ indicated as 1 and 2 in PS shown in
sketch. The difference in ground state energy for these donors obtained from
magnetic field positions of corresponding lines ($\Delta H\sim 0.6$ $kOe$)
is about $0.1$ $meV$. As seen from fig.4 the relative population of donors 1
by electrons- $(S_1/S_2)$ increases with increasing of electric field about $%
10\%$.

The growth technology for LP $n-GaAs$ films on $p-GaAs$ bulk samples
requires substantially higher temperature than for the GP samples. This is
the reason why LP samples in comparison with the GP ones have more
inhomogeneous SI distribution.

\section{CONCLUSION}

The new broadening mechanism of $1s\rightarrow 2p_{+1}$ SIPS lines were
established which takes place in n-GaAs samples with $N_D>10^{14}cm^{-3}$
and connected with fluctuations of potential arising as a result of SI
inhomogeneous distribution in samples. At $E<E_{bd}$ homogeneous and
inhomogeneous SI distribution in samples causes the different dependence on
external electric field for line width. Results obtained extends the
opportunity of n-GaAs samples quality control by SIPS method. In particular,
they allows to check more homogeneous sample among having nearly equal $N_D$
and $K$.

\section{ACKNOWLEDGMENTS}

Samples used in this work with their $N_D$, $K$ and $\mu _{77}$ parameters
are from different sources; samples 2-5 from Dr.T.S.Lagunova, 1 and 6 from
Dr.Y. V. Ivanov, high purity sample7 from Dr. A.Osutin. I would like to
acknowledge them. I also thank prof. V.I. Ivanov-Omskii for supports when I
carried out these experiments at his laboratory.

\begin{table}
  \begin{tabular}{|l|l|l|l|l|l|l|} \hline
     {Samples}  &  {$N_D \cdot 10^{-14}$}  &  {$k=N_A/N_D$}  &  {$N_i \cdot 10^{-14}$}  &  {$\mu _{77}\cdot 10^{-4}$}  &  {$\Delta _{\frac 12}$ $(\mu $$eV)$}  &  {$\Delta _{\frac 12}$ $(\mu $$eV)$} \\ 
     {}  & {$cm^{-3}$}  &  {}  &  {$cm^{-3}$}  &  {$cm^2/V \cdot s$}  &  {$H\approx 36kOe$}  &  {$H\approx 61kOe$} \\ \hline
     1GP       &  4.3       &  0.52       &  4.5       &  10       &  64       &  70  \\
     2CP       &  7.3       &  0.65       &  9.6       &  7.3       &  70       &  80  \\
     3LP       &  8.7       &  0.53       &  9.0       &  7.3       &  150       &  160  \\
     4GP       &  15.0       &  0.38       &  11.0       &  5.7       &  85       &  90  \\
     5LP       &  4.22       &  0.36       &  3.0       &  8.6       &  136       &  140  \\
     6GP       &  4.3       &  0.52       &  4.5       &  10       &  62       &  65  \\
     7GP       &  high purity       &-       &  -       &  -       &  12       &  8  \\ \hline
  \end{tabular}
  \caption{Parameters and widths of $n-GaAs$ $1s\rightarrow 2p_{+1}$ donors lines at linear region of CVC\label{key}}
\end{table}

\newpage\ 

CAPTIONS

Fig.1: (fig1.gif)

PC as a function of magnetic field for sample1 (lower PS) and sample3 (upper
PS) at $\lambda \approx 119\mu $ (left) and $\lambda \approx 84\mu $ (right)
wavelengths radiation. Applied voltage is about $1V$.

Fig.2: (fig2.gif)

PS of different donors of high quality sample7 ($\lambda \approx 119\mu $).
The line width is not depends on applied voltage, as in the case of [8].

Fig.3: (fig3.gif)

Electric field dependence of $1s\rightarrow 2p_{+1}$ line width of two
samples ($\lambda \approx 119\mu $): 1 - sample1; 2 - sample5; 3 - CVC of
sample1 at $E\perp H$ and $H=36.5$ $kOe$. Sample 5 has nearly the same CVC.
Distance between contacts is about $3$ $mm$.

Fig 4: (fig4.gif)

The electric field dependence of two shallow donors $1s\rightarrow 2p_{+1}$
lines intensity relation $S_1/S_2$,which ground state energy difference is $%
\sim $ 0.1 meV .The PS is shown in sketch (sample ,6).

\end{document}